\documentclass[aps,12pt]{revtex4}
\usepackage{graphicx}
\def\bd{\begin{displaymath}}
\def\be{\begin{equation}}
\def\ed{\end{displaymath}}
\def\ee{\end{equation}}

\begin{document}
\title{\bf On the Positron Fraction and the Spectrum of the Electronic Component in Cosmic Rays}
\author{R. Cowsik and B. Burch\\
\emph{Physics Department and McDonnell Center for the Space Sciences\\
Washington University, St. Louis, MO 63130}}

\begin{abstract}
The recent observations of the positron fraction in cosmic rays by PAMELA indicate that the fraction of positrons to the total electronic component in cosmic rays initially decreases in the energy region 1-10 GeV and increases thereafter. In this paper, we show that it is natural to expect such an increase of the positron fraction within the context of cosmic ray propagation  models. It is shown that this ratio should reach an asymptotic value of $\sim$0.6 at very high energies. The specific measurements by PAMELA help us to distinguish amongst various models for cosmic ray propagation, and in particular, they support the nested leaky box model. They also provide, in conjunction with the observations of the total electronic component by HESS, FERMI, ATIC, and other experiments, a way of  estimating the spectrum of electrons directly accelerated by discrete sources of cosmic rays in the Galaxy.
\end{abstract}
\maketitle
\section{Introduction}
\begin{figure}
 \includegraphics[width=12cm]{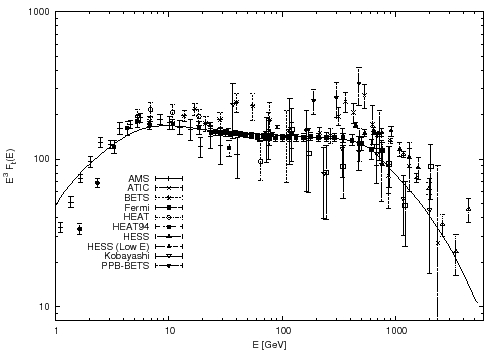}
  \caption{Shown here is the total electron spectrum as measured by HESS, FERMI, ATIC, and other experiments. A smooth fit used for computational purposes is also shown.}\label{fig:HESS}
\end{figure}

Direct observation of the cosmic ray electronic component (which includes both electrons and positrons with no charge discrimination) dates back to the early 1960s, and since that time, the energy range and the sensitivity of the observations have increased systematically. To date, we have at hand data from three new instruments, FERMI \cite{FERMI}, HESS \cite{HESS}\cite{HESSLow}, and ATIC \cite{ATIC}, that have the requisite sensitivity to measure, with good statistical accuracy, the spectrum of the total electronic component ($e^-+e^+$) well into the TeV region. The reported spectrum in the region $E\geq$10 GeV is parameterized as
\be\label{eq:spect} f_t(E)=\kappa E^{-\Gamma} \ee
with $\Gamma=3.05$ up to $\sim$ 1 TeV, and the value of $\Gamma$ increases to $\sim$ 3.9 at higher energies. We reproduce in fig. \ref{fig:HESS} their observations and a compilation of the results of other measurements. The spectral slope below 10 GeV progressively flattens to a slope of $\sim$1.7. Uncertainties in the fluxes are introduced due to solar modulation effects below an energy of a few GeV. We also show, in the same figure, a smooth fit to all the data that we adopt for some of the calculations.

The electronic component in cosmic rays, because of its interactions with radiation fields such as starlight, the microwave background, and magnetic fields in the galaxy, has been particularly useful in understanding the origins and propagation of energetic particles in the Galaxy \cite{Cowsik66}. This and other early considerations of the effects on the spectral shape of the cosmic ray electrons were carried out in the context of a smooth distribution of cosmic ray sources in the galaxy, and the transport was described within the framework of the leaky box model \cite{Cowsik67}. There were also attempts to calculate the flux of nuclear secondaries like Li, Be, and B, as well as positrons, in cosmic rays within the context of a nested leaky box model which took into consideration the effects of storage of cosmic rays in a small bubble surrounding the compact sources of cosmic rays \cite{Cowsik73}\cite{Cowsik75}. The key consideration for this model was the anisotropy of cosmic rays at high energies. If the residence time of cosmic rays in the Galaxy reduced with energy to accommodate the decreasing ratio of secondary to primary nuclei in cosmic rays, then the expected anisotropy would increase correspondingly at high energies.

\begin{figure}
 \includegraphics[width=12cm]{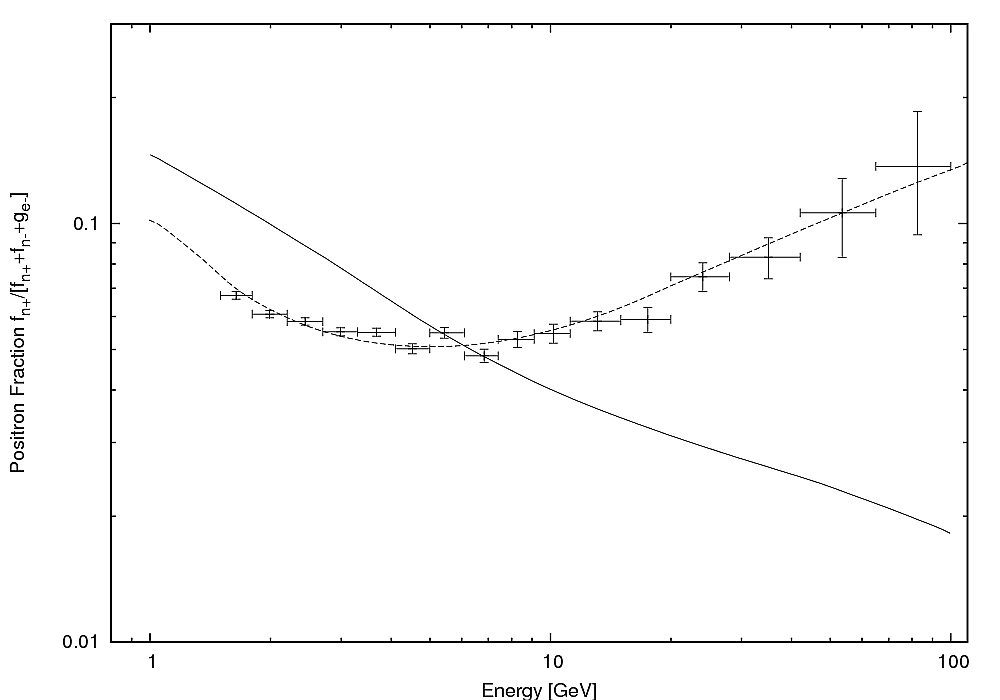}
  \caption{The positron fraction as measured by PAMELA (+'s) is shown here. Also included is the calculation by Moskalenko and Strong\cite{Moskalenko} (solid line) and a smooth fit through the data (dashed line).}\label{fig:PAMELA}
\end{figure}

The recent observations of positrons in cosmic rays by PAMELA \cite{PAMELA} has created much excitement because of the possible connection of these observations with annihilation of dark matter in the Galaxy. Their observations of the ratio $R$, of positrons to the total electronic component in cosmic rays is reproduced in fig. \ref{fig:PAMELA}. The positron fraction at $\sim$1.64 GeV was measured to be $\sim$0.0673, which decreases to $\sim$0.0483 at $\sim$6.83 GeV and thereupon increases monotonically, reaching a value of 0.137 at a mean energy of 82.55 GeV. It is this monotonic increase that is being called anomalous, as it does not conform to the prediction of the currently popular model of the cosmic ray propagation \cite{Moskalenko}\cite{Strong}. Accordingly, this paper begins with a review of the cosmic ray propagation models. The PAMELA measurements are to be viewed in the context of the measurements of the total electronic component and a compilation of the available data as shown in fig. \ref{fig:HESS}.

The cosmic ray nucleonic component is dominated by protons with some neutrons coming in bound as He and other nuclei. The nucleon spectrum may be represented as
\be f_n(E)=\kappa_nE^{-\beta}\ee
where $\beta\approx2.6-2.7$ (we adopt $\beta=2.65$) in the energy region 1 GeV/nucleon to $\sim$10$^6$ GeV/nucleon beyond which the slope may increase to $\sim$3. The $p/n$ ratio effectively determines the $e^+/e^-$ ratio generated by cosmic rays through interactions with matter in the sources and the interstellar medium through which they propagate before they leak out of the Galaxy. The theoretical calculation of $e^+/e^-$ generated through nuclear interaction of cosmic ray nuclei yield $\sim$2 \cite{Protheroe}. On the other hand, the observations of the $\mu^+/\mu^-$ ratio in cosmic rays gives $\mu^+/\mu^-\approx$1.3 \cite{Hayakawa}. We show our results for both of these values for the secondary positrons and electrons.

In section II, we provide a brief overview of the models for cosmic ray propagation, and in section III, we discuss the PAMELA results, first in a model independent way, and then compare it with the expectations of the various models.
Finally, section IV is devoted to a discussion of the main results of this paper and related matters.

It should be noted that much work as has been done recently to explain the PAMELA data in terms of dark matter, pulsars, supernova remnants, and other astrophysical objects. Instead of listing the many papers which address these issues, we refer the reader to \cite{Profumo} which contains references to many recent works.


\section{Brief Overview of Models of Cosmic Ray Propagation}
It is generally accepted that cosmic rays are accelerated in a large number of discrete sources distributed in the Galaxy, and the cosmic rays propagate from these sources moving along randomly oriented trajectories, akin to diffusion, until they leak away from the Galaxy. During such a propagation, the cosmic rays might interact with the interstellar matter, the radiation, and magnetic fields. Any secondaries generated through such interactions, if charged, will be confined by the interstellar magnetic fields and will therefore follow the same kind of random paths as the primaries before escaping from the Galaxy. The various propagation models are characterized by the specific form chosen for the ``vacuum path length distribution'' \cite{Cowsik67} which describes the probability $P(t)$ that the cosmic rays spend in any given region, such as a cocoon surrounding the sources, or in the general interstellar medium before escaping into the intergalactic space. The term ``vacuum'' emphasizes the fact that in specifying $P(t)$, one considers hypothetical particles which do not suffer interactions or lose energy during propagation. The effects of these processes are to be added later on.

\subsection{The Leaky Box Model}
\begin{figure}
 \includegraphics[width=12cm]{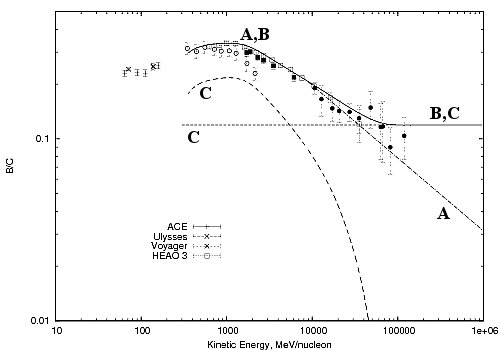}
  \caption{The observed B/C secondary to primary ratio is plotted (points from a compilation in \cite{Strong}) along with the power law extrapolation at high energies (dot-dashed line, Model A), a constant extrapolation (solid line, Model B), and a two-component fit (dotted lines, Model C).}\label{fig:BC}
\end{figure}

\begin{figure}
 \includegraphics[width=12cm]{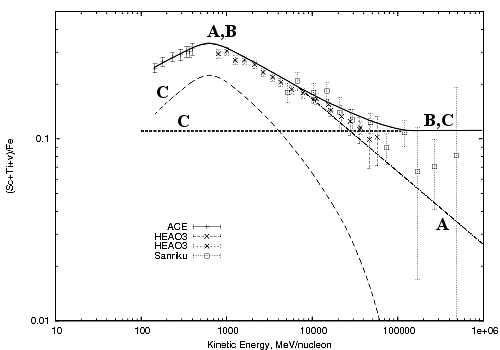}
  \caption{The observed (Sc+Ti+V)/Fe secondary to primary ratio is plotted (points from a compilation in \cite{Strong}) along with the power law extrapolation at high energies (dot-dashed line, Model A), a constant extrapolation (solid line, Model B), and a two-component fit (dotted lines, Model C).}\label{fig:SC}
\end{figure}

In its simplest original form \cite{Cowsik67}, one assumes that $P(t)$ has a broad distribution with significant amplitude near $t=0$, exemplified by a simple exponential function
\be P(t)=e^{-t/\tau} \ee
where $\tau$ is called the escape lifetime of the cosmic rays. In the original version, $\tau$ was assumed to be sensibly independent of energy beyond $\sim$1-2 GeV. Thereafter, since the discovery that the ratio of the fluxes of secondary cosmic ray nuclei to those of the primaries was a decreasing function of energy, $\tau$ was considered to decrease with energy to accommodate the observations. We summarize in figs. \ref{fig:BC} and \ref{fig:SC} the available observations. The crucial issue here is how $\tau$ behaves at energies beyond 10-20 GeV where the observations have low statistical significance or are non-existent at much higher energies. Most conventional models today \cite{Moskalenko}\cite{Strong} assume that
\be\label{eq:TauA}\tau_A(E)\sim\tau_0E^{-\Delta}\quad for~E>2GeV/n~~(Model~A)\ee
with $\Delta\approx0.4-0.5$. Such an extrapolation to high energies is shown in a dot-dashed line in figs. \ref{fig:BC} and \ref{fig:SC}.

Alternatively, we may assume that $\tau(E)$ becomes nearly constant at high energies
\be\left.\begin{array}{ll}
   \tau_B(E)\sim\tau_A(E) & \quad for~E<10GeV\\
    \tau_B(E)\sim\tau_G\sim~constant & \quad for~E\gg10GeV
    \end{array}\right\}~~(Model~B)\ee
This is shown as a solid line in figs. \ref{fig:BC} and \ref{fig:SC}. The latter form of $\tau\sim\tau_G$ at high energies predicts a lower level of anisotropy of cosmic rays compared to an ever decreasing $\tau_A(E)$. We will refer to the two models described briefly here as leaky box model A and leaky box model B respectively.

\subsection{The Nested Leaky Box Model}
An alternate way of accommodating the falling secondary to primary ratio is in the context of the nested leaky box model \cite{Cowsik75}. Here, one assumes that subsequent to acceleration, cosmic rays spend some time in a cocoon-like region surrounding the sources, interacting with the matter there to generate some of the secondaries at lower energies. The residence time $\tau_s$ in the source region is energy dependant, decreasing with increasing energy. On the other hand, once these cosmic rays enter the general interstellar medium, their subsequent transport becomes independent of energy and the residence time becomes equal to $\tau_G$.
\be\left.\begin{array}{ll}
    \tau_s(E)\sim\tau_B(E)-\tau_G & \quad for~1~GeV<E<10~GeV\\
    \tau_G(E)\sim~constant & \quad for~1~GeV<E<10^6~GeV
    \end{array}\right\}~(Model~C).\ee
The net effect of the interactions in these two regimes is to generate the correct ratio of the fluxes of secondary nuclei to those of their primaries.

In this model, the anisotropies of cosmic ray fluxes remain constant and do not increase with energy. Moreover, the spectrum of cosmic ray primaries, say the nuclear component, is more easily understood. To see this, let $S_n(E)$ represent the injection rate of cosmic rays into the Galactic volume per unit volume and unit time at energy $E$ per unit energy interval:
\be S_n(E)\sim E^{-\alpha}.\ee
The spectrum $f_n(E)$ expected in the Galaxy is given by
\be f_n(E)\sim\tau_GS_n(E)\sim\tau_GE^{-\alpha}.\ee
On the other hand, in the simple leaky box model A,
\begin{eqnarray}
     f_n(E)&\sim&S_n(E)\tau_A(E)\nonumber\\
     &\sim&E^{-(\alpha+\Delta)}\quad~~~~ for~E>2~GeV/n
\end{eqnarray}
which will fit the observations for the choice $\alpha\approx2.2-2.3$ with the sources accelerating a flatter spectrum than the one that is observed. This flatter spectrum must continue up to very high energies, up to which the residence time continues to decrease as $E^{-\Delta}$. If such a rapid decrease of residence time stops at any energy and becomes constant, then $f_n(E)$ will display $E^{-\alpha}$ behavior at higher energies, or until $S_n(E)$ itself changes its slope.

In the leaky-box model B, the source function $S_n(E)$ should have an index $\alpha$=2.2 at $E<10$ GeV/n, which changes to $\alpha=2.65$ at $E\gtrsim10-20$ GeV, coincident with the change in behavior of $\tau_B(E)$, compensating its change and generating a smooth power law for $f_n(E)$. Thus, in the nested leaky box model, the observed spectral slopes simply correspond to that generated in the acceleration process in the cosmic ray sources. Also, with constant $\tau_G$, the expected anisotropies do not increase with increasing energies, but remain sensibly constant.

\subsection{Spectrum of Secondary Electrons and Positrons in Cosmic Ray Propagation Models}
The generation of electrons and positrons in the interactions of the cosmic ray nuclear component occurs through the production of mesons, mainly pions, which decay to muons which in turn decay into electrons or positrons, transferring, on the average, a fraction of about 0.05 of the energy per nucleon of the primary. This is in contrast with the production of secondary nuclei, such as boron from the collision of carbon nuclei, where boron emerges with almost the same energy per nucleon as the primary carbon nucleus. This difference in their production characteristics leads to nearly identical source spectra $S_{n^-}$ and $S_{n^+}$ for the secondary electrons and positrons $\sim E^{-\beta}$ in all the three models: A, B, and C.

On the other hand, their equilibrium spectra $f_{n^+}(E)$ and $f_{n^-}(E)$ are markedly different in the three models. At energies where the energy losses due to synchrotron radiation and inverse Compton scattering on radiation fields are not important. The three models generate the spectra noted below:
\begin{eqnarray}\label{eq:fn}
f_{n^+}&\sim& S_{n^+}(E)\tau_A(E)\sim\tau_0E^{-(\beta+0.4)}\quad(Model~A)\\
&\sim&S_{n^+}(E)\tau_B(E)\sim\tau_0E^{-(\beta+0.4)}\quad for~E<10~GeV\nonumber~(Model~B)\\
&\sim&S_{n^+}(E)\tau_G\sim E^{-\beta}\quad\quad~~~~~~~~~~ for~E>10~GeV\nonumber~(Model~B)\\
&or&~S_{n^+}(E)\tau_G\sim E^{-\beta}\quad ~~~~~~~~~~~~for E>1~GeV~(Model~C)\nonumber
\end{eqnarray}
The spectra for the secondary electrons are similar to those given in eq. \ref{eq:fn} except that because of the dominance of the protons in the cosmic ray beam, the production rate of positions is higher, with
\be \frac{S_{n^-}(E)}{S_{n^+}(E)}=\eta.\ee
This ratio $\eta$ is theoretically estimated from the characteristics of high energy interactions to be $\sim$0.5 \cite{Moskalenko}; on the other hand, the direct observation of the $\mu^-/\mu^+$ ratio indicates a value of $\sim0.8$ \cite{Hayakawa}. In either case, $\eta$ is essentially independent of energy beyond a few GeV.

Thus we see that, at high energies (E$\gg$10 Gev), in the leaky-box model B and in the nested leaky box model C, the secondary positron and electron spectra are power laws with indices $\beta$ equal to that of the spectrum of the nuclear component in cosmic rays. At very high energies, the energy losses due to synchrotron emission and inverse Compton scattering will steepen these spectra to $f_{n^+}=E^{-(\beta+1)}$. 

\section{Analysis of the Positron Fraction Observed by PAMELA}
\begin{figure}
 \includegraphics[width=12cm]{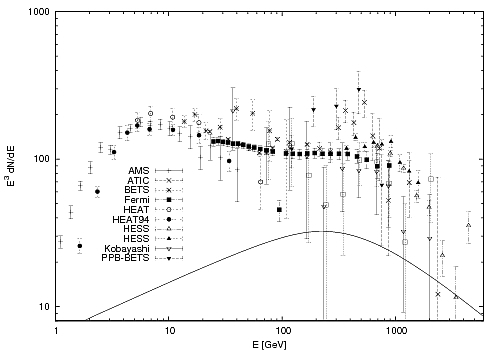}
  \caption{Here we have subtracted $f_{n^+}(E)$ and $f_{n^-}(E)$ from the total spectrum of the electronic component and have shown $g_{e}(E)$, the spectrum of electrons generated by the cosmic ray sources. The primary electron component $g_{e}(E)$ is plotted for the HESS, FERMI, ATIC and other data using $\eta=0.45$. The positron spectrum from the nested leaky box model C is plotted as well (solid line).}\label{fig:PRIMARY}
\end{figure}
\begin{figure}
 \includegraphics[width=12cm]{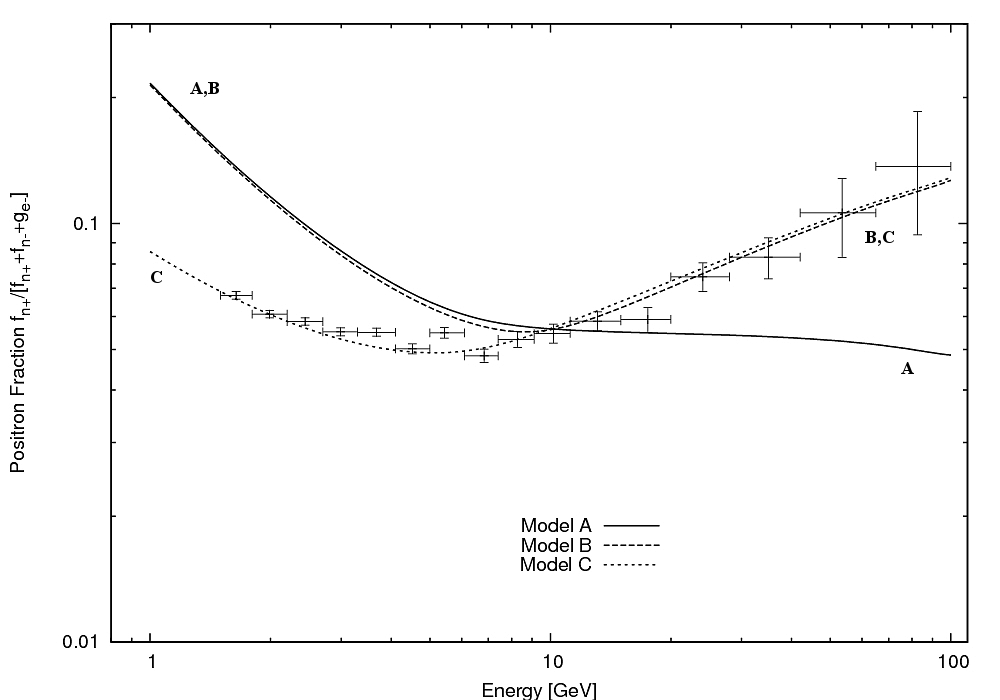}
  \caption{The theoretically calculated positron fraction in models A (similar to that of \cite{Moskalenko} as shown in fig. \ref{fig:PAMELA}), B, and C are compared with the observations. All calculations are normalized at $\sim10$ GeV}\label{fig:PAMMODELS}
\end{figure}
We find it useful to write the observed positron fraction $R(E)$ in terms of the various components:
\be R(E)=\frac{f_{n^+}}{f_{n^+}+f_{n^-}+g_{e^-}}.\ee
Here, $f_{n^+}$ and $f_{n^-}$ represent the positron and electron spectra generated as secondaries of the nuclear component of cosmic rays, and $g_{e^-}$ is the spectrum of electrons resulting from direct acceleration in the sources. Note that there is no direct contribution to positrons from the source. It is convenient sometimes to work with the inverse of $R(E)$ given by
\be P(E)=\frac{1}{R(E)}=\frac{f_{n^+}+f_{n^-}+g_{e^-}}{f_{n^+}}= 1+\eta+\frac{g_{e^-}}{f_{n^+}}.\ee
This allows one to find the spectrum of electrons generated by the sources $g_{e^-}$ as
\be g_{e^-}(E)=[P(E)-(1+\eta)]f_{n^+}(E).\ee
This spectrum $g_e(E)$, generated in the Galaxy exclusively by the cosmic ray sources, is shown in fig. \ref{fig:PRIMARY}.

Alternatively, we may just assume the functional form for $f_{n^+}(E)$ given by the various propagation models and calculate the positron fraction by dividing this by $f_t(E)$, the total spectrum of electrons measured by FERMI, HESS, and other experiments,
\be R_M(E)=\frac{f_{n^+}(E)}{f_t(E)}\ee
which is shown in fig. \ref{fig:PAMMODELS} along with the data from PAMELA. The normalization of the theoretical curves is such as to provide the best possible fit to the three models A, B, and C described earlier. (This normalization may indeed be explicitly calculated as it is proportional to $\tau_A(E)$, $\tau_B(E)$, and $\tau_G$ respectively for the three models and depends on the density of matter in the propagation region, the spectral flux of the nuclear component, the cross section for meson production, decay kinematics, etc.) In depicting the three model curves, we have included the effect of the energy losses at high energies.

Comparison of the theoretical expectations of the positron fraction with the PAMELA data indicates that model A provides a rather poor fit to the observation, as already noted by several authors \cite{PAMELA}. A careful calculation of the positron fraction under the general assumptions of model A was carried out over a decade ago by Moskalenko and Strong \cite{Moskalenko}. Our estimates here are essentially the same as that derived by them. Even though both model B and model C predict nearly identical injection spectra, the equilibrium spectra at low energies differ drastically with each other. In model B, the injection spectrum has to be multiplied by $\tau_B(E)$ to get the equilibrium spectrum. On the other hand, in the nested leaky box model, we need to multiply only by $\tau_G$ \cite{Cowsik73}\cite{Cowsik75}. Both these models predict identical equilibrium spectra at high energies, for $E\gg10$ GeV. The good fit to observed the positron fraction shows that the residence time of cosmic rays is essentially independent of energy for $E>10~GeV$, a constancy that is expected to continue up to $\sim10^5~GeV$.

In choosing between model B and model C, the latter is preferred from considerations of the spectra of primary nuclei as well. This is because for a simple power law input from the sources having a form $S_n(E)\sim E^{\beta}$, model B would be expected to yield a spectral form $f_n(E)\sim E^{-(\beta+\Delta)}$ below $\sim10$ GeV and $E^{-\beta}$ at higher energies. On the other hand, the observed spectra of all the nucleonic components are simple power laws of slope $\sim E^{-2.65}$ with no changes of slope in the tens of GeV region. Thus we conclude that the nested leaky box model provides good fit with the PAMELA data and is preferred also from consideration of other cosmic ray observations.


\section{Discussion and Related Matters}
The main result that emerges from the present analysis is that the nested leaky box model provides a good fit to the positron fraction observed by PAMELA. The model is also consistent with other observations of cosmic rays. Until good measurements of the positron fraction was available, there was no easy means of choosing amongst the various models. The fact that the nuclear secondaries, such as Li, Be, and B, emerge from nuclear interactions with essentially the same energy per nucleon as their parents, C, N, and O, was the main cause for this uncertainty. However, the fact that the positrons carry, on the average, a fraction of only about 0.05 of the energy of their nuclear primaries breaks this degeneracy, allowing a choice to be made. Improvements of the measurements of the spectra of both the secondary nuclei and of positrons will help in fixing, more firmly, the parameters of the nested leaky box model.

The recent measurements of the electronic component of cosmic rays by the PAMELA, HESS, ATIC, and FERMI groups have opened up many interesting new avenues to discuss cosmic ray propagation. Some of these were discussed or hinted at in this paper.

\end{document}